\documentclass [12pt]{article}

\usepackage {hyperref}

\hypersetup{urlbordercolor=0 0 0,pdfborder=0 0 1 [3 2]}

\begin{document}

\title{Peculiar Fields in Maxwell's Equations.}
\author{A. N. Ageev * and A. G. Chirkov **
* (Ioffe Physicotechnical Institute, Russian Academy of Sciences, St. Petersburg, Russia)
** (St. Petersburg State Technical University, St. Petersburg,
Russia )}
\date{\today}
\maketitle

Peculiar Fields in Maxwell's Equations.

A. N. Ageev * and A. G. Chirkov **

* (Ioffe Physicotechnical Institute, Russian Academy of Sciences, St.
Petersburg, Russia)

** (St. Petersburg State Technical University, St. Petersburg, Russia )

Abstract---A theoretical analysis of the excitation of an infinitely long
solenoid by oscillating current has revealed the existence of specific
potentials in the space outside the solenoid, which can affect electron
diffraction in an experiment similar to the Aharonov-Bohm effect. Thus,
these time-dependent potentials are physical fields possessing a number of
specific features, which set them off from the fields known heretofore.


The peculiar phenomenon, predicted in 1939 and 1949 [1, 2] and rediscovered
and studied theoretically in considerable detail in 1959 [3], was
subsequently called the Aharonov-Bohm (AB) effect. It consists essentially
in that in propagating through a region with no magnetic or electric field
present, but where the vector or scalar potential is nonzero, the de Broglie
wave corresponding to a quantum charged particle is acted upon by the
latter. These conditions are best realized in a static regime, which was
exactly the case studied before the 1990s. While a long discussion has
certainly contributed to a proper understanding of the AB effect (see, e.g.,
reviews [4]), heated debates on this issue are still continuing in the
literature. Most clear and sequentially theory of effect AB enunciated in
paper of prof. D. H. Kobe [5].

Based on the totality of the experiments performed, one has to admit that
the AB effect can exist only if there are potentials, which do not generate
fields and cannot be removed by gauge transformation. We have termed them
``zero-field potentials''. Note that zero-field potentials, which transform
only the phase of a wave function, are responsible for the AB effect in all
the papers published heretofore and dealing with the static case. In a
general case, such potentials satisfy the relations

 $ - \,c^{ - 1}\partial {\rm {\bf A}}^{0} / \partial t\, - \,grad\,\varphi
^{0}\,\,\, = \,\,0$ and $\,rot\,{\rm {\bf A}}^{0}\, = \,\,0,$ (1)

\noindent
where the upper indices of the potentials refer to the zero-field
potentials. Because such potentials should obviously have the form

\begin{equation}
\label{eq1}
\,{\rm {\bf A}}^{0} = \,\,\,grad\,\chi {\rm è}
\quad
\varphi ^{0}\, = \, - \,{\frac{{1}}{{c}}}\,{\frac{{\partial \,\chi
}}{{\partial \,t}}},
\end{equation}

\noindent
the $\,\chi $ function was erroneously identified in practically all
publications with the gradient potential transformation function, and this
is what gives rise frequently to misunderstanding.

After the convincing experiments of Tonomura \textit{et al}. [6], the possibilities of
studying the static AB effect at the present level of technology were
apparently exhausted, and the researchers turned their attention to the
investigation of the time-dependent, or quasi-AB effect [7]. However, in
this work, which has certainly produced fruitful results, the potentials
responsible for the quasi-AB effect were introduced artificially, without
discussing in any way their nature. Nevertheless, the origin of these
potentials (fields) is a major issue in the separation of the AB effect from
the general variation of the de Broglie wave-interference pattern due to
Lorenz force.

We maintain that in the regions of space with no currents present the total
potentials can be presented, generally speaking, in the form

 ${\rm {\bf A}} = \,\,{\rm {\bf A}}^{f} + \,{\rm {\bf A}}^{0}$ and $\varphi \, =
\,\varphi ^{f}\, + \,\varphi ^{0}$, (3)

\noindent
where index $f$ refers to ``field'' potentials corresponding to nonzero
electromagnetic fields:

\begin{equation}
\label{eq2}
{\rm {\bf E}} = \, - \,c^{ - 1}\partial {\rm {\bf A}}^{f} / \partial t\, -
\,grad\,\varphi ^{f}\,,
\quad
\,
{\rm {\bf B}} = \,rot\,{\rm {\bf A}}^{f}.
\end{equation}

Index $0$ in Eq. (3) identifies zero-field or excess potentials defined by
relations (1). Note that the ``excess'' potentials have been long in use in
mathematical physics [8]; they are necessary when solving Maxwell's
equations with boundary conditions.

We are going to demonstrate the above in a specific example. Consider
circular currents flowing in a region of space to form an infinitely long
cylinder of radius $R$ (a solenoid with circular currents). Choose a
cylindrical reference frame ($\rho ,\,\alpha ,\,z$) with the axis $z$
coinciding with the solenoid axis. In the magnetostatic case, the solution
within the infinite solenoid ($\,0 \le \,\rho \, < \,R$) can be chosen in
the form $A\,_{1\alpha } \,\, = \,\,c_{1} \,\rho \,\,\,\,\,(\,A_{1}^{f} =
A\,_{1\alpha } ,\,\,\,\,A_{1}^{0} \, = \,0)$. In the outer region ($\,\rho
\, > \,R$), the solution has the form $A\,_{2\alpha } \,\, = \,\,c_{2} \, /
\,\rho \,\, + \,c_{3} \,\rho $. The system being infinite, one cannot
require the potential to vanish at large distances. As is clear from purely
physical considerations, the magnetic field outside the solenoid is zero,
i.e., $A_{2}^{f} = \,\,0$. Therefore, the only potential that can exist in
the outer region is ${\rm {\bf A}}^{0}$, which satisfies the additional
condition $\,rot\,{\rm {\bf A}}^{0}\, = \,\,0,$ and it is this condition
that identifies the correct solution $A\,_{2\alpha } \,\, = \,\,c_{2} \, /
\,\rho \,\,$. The potential in the outer region is essentially the
zero-field potential, so that ${\rm {\bf A}}_{2} \, = \,grad\,\chi $, but
because this region is doubly connected, the $\chi $ function is
multivalued, and ${\oint\limits_{L} {{\rm {\bf A}}_{2} \,d{\rm {\bf l}}}
}\,\, \ne \,\,0$, despite the fact that in this region ($R\, < \,\rho \, <
\,\infty $)$\,rot\,{\rm {\bf A}}_{2} \, \equiv \,\,0$.

The above separation of the potentials into the field and zero-field ones
permits one to find the zero-field potentials for a time-dependent current
as well. As before, we assume that circular currents flow in a region of
space to form an infinitely long cylinder. The reference frame will be left
unchanged. The current can be described by the following relations

 $j_{\alpha } $ ($\rho ,\,\alpha ,\,z$) $ = \,\,I_{0} \,\delta \,(\rho \, -
\,R)\,\exp \,i\,\omega t\,$, $j_{\rho } \, = \,j_{z} \, = \,0$, (5)

\noindent
where $R$\textit{ }is the solenoid radius, $\omega $ is the cyclic frequency of the
current, and $I_{0} \, = \,J\, / \,2\pi R$; here $J$ is the current in the
cylinder wall per unit length of the solenoid.

The nonzero vector-potential components $A_{\rho } $ and $A_{\alpha } $can
be written [9]

\begin{equation}
\label{eq3}
A_{\rho } \, = \,{\int\limits_{V} {j_{\alpha } \,(} }{\rm {\bf {\rho
}'}})\,\sin \,(\alpha \, - \,{\alpha }')\,G\,({\rm {\bf \rho }},\,{\rm {\bf
{\rho }'}})\,d{V}',
\end{equation}

\begin{equation}
\label{eq4}
A_{\alpha } \, = \,{\int\limits_{V} {j_{\alpha } \,(} }{\rm {\bf {\rho
}'}})\,\cos \,(\alpha \, - \,{\alpha }')\,G\,({\rm {\bf \rho }},\,{\rm {\bf
{\rho }'}})\,d{V}',
\end{equation}

\noindent
where $G\,({\rm {\bf \rho }},\,{\rm {\bf {\rho }'}})\,\, = \,\, -
\,{\frac{{i\pi }}{{c}}}\,H_{0}^{\left( {2} \right)} \,(k{\left| {\,{\rm {\bf
\rho }} - \,{\rm {\bf {\rho }'}}} \right|})$ is the Green function of the
Helmholtz equation [9], $H_{0}^{\left( {2} \right)} $ is the Hankel
function, $k\, = \,\omega / \,c$, and $d\,{V}'\,\, = \,{\rho }'\,\,d{\rho
}'\,d\,{\alpha }'$. Here and in what follows, the harmonic dependence on
time is omitted. The integrals entering Eq. (6) can be easily calculated
using the rules of the totals for the Hankel functions [9]

\[
H_{0}^{\left( {2} \right)} \,(k\,\sqrt {\rho ^{2}\, + \,R^{2}\, - \,2\rho
R\cos \,(\alpha - {\alpha }')} )\, = \,
\]

\[
 = \,{\sum\limits_{m = - \infty }^{\infty } {e^{ - im(\alpha - \alpha '\,)}}
}
{\left\{ {{\begin{array}{*{20}c}
 {H_{m}^{\left( {2} \right)} \,(kR)\,\,J_{m} \,(k\rho ),\,\,\rho < \,R}
\hfill \\
 {J_{m} \,(kR)\,H_{m}^{\left( {2} \right)} \,(k\rho ),\,\,\rho \, > \,R}
\hfill \\
\end{array} }} \right.}.
\]

As a result, we obtain

 $A_{\alpha } \, = \, - \,{\frac{{2i\pi ^{2}\,I_{0} \,R}}{{c}}}
{\left\{ {{\begin{array}{*{20}c}
 {H_{1}^{\left( {2} \right)} (kR)\,J_{1} \,(k\rho ),\,\rho < R} \hfill \\
 {J_{1} \,(kR)H_{1}^{\left( {2} \right)} (k\rho ),\,\rho > R} \hfill \\
\end{array} }} \right.}$ and $A_{\rho } \, = \,0$. (7)

In the static case ($\omega \, \to \,0$), one obtains from these relations
the well-known expressions

 $A_{\alpha } \, = \,J\,\rho \, / \,c\,R\,\,(\rho \, < \,R)$ and $A_{\alpha } \,
= \,J\,R\, / \,c\,\rho \,\,(\rho \, > \,R)$, (8)

Consider in more detail the potential of Eq. (7) in the outer region, which
is of major interest for us here

\[
A_{\alpha } \, = \,Q\,H_{1}^{\left( {2} \right)} (k\rho )\, \equiv \,Q
[J_{1} \,(k\rho )\, - \,\,i\,Y_{1} \,(k\rho )]
 =
\]

 $ = \,\,\,Q\,\,$ {\{}$\,{\frac{{2i}}{{\pi k\rho }}}\, + \,{\left[ {1 -
{\frac{{2iC}}{{\pi }}} - \,{\frac{{2i}}{{\pi }}}\ln \left( {{\frac{{k\rho
}}{{2}}}} \right)} \right]}{\sum\limits_{m = 0}^{\infty } {{\frac{{( -
1)^{m}}}{{m!\,\Gamma \,(m + 2)}}}\left( {{\frac{{k\rho }}{{2}}}} \right)^{2m
+ 1}\,} }\,\, + \,$

 $ + {\frac{{i}}{{\pi }}}{\sum\limits_{m = 0}^{\infty } {{\frac{{( -
1)^{m}}}{{m!\,(m + 1)!}}}\left( {{\frac{{k\rho }}{{2}}}} \right)^{2m + 1}}
}{\left[ {{\sum\limits_{j = 1}^{m} {{\frac{{1}}{{j}}}} }\, +
\,{\sum\limits_{j = 1}^{m + 1} {{\frac{{1}}{{j}}}} }} \right]}$ {\}}, (9)

\noindent
where $Q\,\, = \,\, - \,{\frac{{2i\pi ^{2}\,I_{0} \,R}}{{c}}}J_{1} \,(kR)$,
$C$ is Euler's constant, and $Y_{1} $ is the Neumann function.

As seen from Eq. (9), the curl of the first term in braces is zero. One can
readily verify that the curls of the other terms in the braces are nonzero.
Thus, in this case the total potential can be separated into the field and
the zero-field potential. As follows from Eq. (1)

\begin{equation}
\label{eq5}
\varphi ^{0}\, =
\quad
{\left\{ {{\begin{array}{*{20}c}

{0,\,\,\,\,\,\,\,\,\,\,\,\,\,\,\,\,\,\,\,\,\,\,\,\,\,\,\,\,\,\,\,\,\,\,\,\,\,\,\,\,\,\,\,\,\,\,\,\rho
< R} \hfill \\
 { - \,{\frac{{4\pi iI_{0} \,R}}{{c}}}\,J_{1} \,(kR)\,\alpha
,\,\,\,\,\,\,\,\rho > R} \hfill \\
\end{array} }} \right.} \quad .
\end{equation}

Separation of the real part of the components of the potentials in Eq. (9)
yields [10]

Re$\,A_{\alpha }^{f} \, = \,W\,\{\pi J_{1} (k\rho )\,\sin \omega t\, -
\,{\left[ {{\frac{{2}}{{k\rho }}}\, + \,\pi Y_{1} (k\rho )} \right]}\,\cos
\omega t\}$, (11a)

Re$A_{\alpha }^{0} \, = \,W\,{\frac{{2}}{{k\rho }}}\,\cos \omega t$, (11b)

\noindent
where $W\, = \,{\frac{{2\pi I_{0} \,R\,J_{1} (kR)}}{{c}}}$.

Consider now the geometry of the Aharonov-Bohm experiment, in which
electrons move around a solenoid along a circle of a given radius. We shall
limit ourselves to the case where the electrons meet on their way nonzero
zero-field potentials, while field potentials are not present. This
situation can be realized by enclosing the solenoid in cylindrical screens,
or, as follows from Eq. (11a), by choosing the trajectory radii of the
electrons and by mathching properly their transit with the current variation
in the solenoid. Substituting now the zero-field potentials in the
Schrodinger equation and using the procedure of the solution proposed in
(Appendices B and D in [7]) but, in contrast to [7], performing time
averaging, we come to the following expression for the intensity of the
interference pattern [11]

\begin{equation}
\label{eq6}
\overline {P} = \,0.5\,\,P_{0} \,\,\,{\left\{ {1 + \,J_{0} \,(S) \cdot \cos
\,[\omega _{e} \,\tau ]} \right\}},
\end{equation}

\noindent
where $S\,\, = \,\,16\,\pi ^{3}\,I_{0} \,\,R\,\mu _{0}^{ - 1} \,\omega ^{ -
1}\,J_{1} \left( {k\,R} \right)$, $\mu \,_{0} \,\,\, =
\,\,{\raise0.7ex\hbox{${ch}$} \!\mathord{\left/ {\vphantom {{ch} {{\left|
{e} \right|}}}}\right.\kern-\nulldelimiterspace}\!\lower0.7ex\hbox{${{\left|
{e} \right|}}$}}\,$ and $J_{0} \,\,$ and $J_{1} $ are the Bessel functions.
For $I_{0} \,\, = \,\,\,158\,\,mA / cm\,\,;\,\,R\,\, = \,\,5\,\mu
m\,;\,\,\omega / 2\pi \,\, < \,\,10^{10}Hz$, we obtain $S$ = 2.45. This means
that the interference pattern should vanish for these parameters. To verify
experimentally this conclusion, one should use preferably electrons in
metallic mesoscopic rings or cylinders [4]

Thus, we believe that the Aharonov-Bohm experiment in both the static and
the time-dependent case is actually an experiment on detection of a field of
a new type in classical electrodynamics. This field has none of the
characteristics inherent in the classical electromagnetic fields, namely,
the energy, the momentum, and the angular momentum. Therefore, these fields
have a high penetration capacity and can be used for information transfer,
with its detection by the AB effect.

REFERENCES

1. W. Franz, Verhandlungen der Deutschen Physikalischen Gesellschaft. 20, 65
(1939).

2. W. Ehrenberg and R. E. Siday, Proc. Phys. Soc. London, Sect. Â 62, 8
(1949).

3. Y. Aharonov and D. Bohm, Phys. Rev. 115, 485 (1959).

4. S. Olariu and I. I. Popesku, Rev. Mod. Phys. 57, 339 (1985); M. Peskin
and A. Tonomura, Lect. Notes Phys. 340, 115 (1989).

5. D. H. Kobe, Annals of Physics 123, 381 (1979).

6. A. Tonomura et al., Phys. Rev. Lett. 56, 792 (1986).

7. B. Lee, E. Yin, Ò. Ê. Gustafson, and R. Chiao, Phys. Rev. A 45, 4319
(1992).

8. A. N. Tikhonov and A. A. Samarskii, Equations of Mathematical Physics
(Nauka, Moscow, 1953; Pergamon, Oxford, 1964).

9. G. T. Markov and A. F. Chaplin, Excitation of Electromagnetic Waves
(Moscow, 1967).

10. A. G. Chirkov and A. N. Ageev, Pis'ma Zh. Tekh. Fiz. 26, 103 (2000)
[Tech. Phys. Lett. 26, 747 (2000)]; A. G. Chirkov and A. N. Ageev, Technical
Physics, 46, 147 (2001).

11. A. G. Chirkov and A. N. Ageev, Solid State Physics 44, 1 (2002).

\end{document}